\documentclass[11pt, twocolumn]{article}

\usepackage[utf8]{inputenc}
\usepackage[T1]{fontenc}
\usepackage{mathptmx}
\usepackage{amsmath, amssymb}
\usepackage{graphicx}
\usepackage{booktabs}
\usepackage{multirow}
\usepackage{xcolor}
\usepackage[margin=2cm]{geometry}
\usepackage{hyperref}
\usepackage{cleveref}
\usepackage{caption}
\usepackage{subcaption}

\hypersetup{colorlinks=true, linkcolor=blue!60!black, citecolor=blue!60!black, urlcolor=blue!60!black}

\graphicspath{{figures/}}

\newcommand{\helmlab}{\textsc{Helmlab}}
\newcommand{\stress}{\textsc{stress}}
\newcommand{\combvd}{\textsc{combvd}}
\newcommand{\ie}{i.e.,\,}
\newcommand{\eg}{e.g.,\,}
\newcommand{\etal}{\textit{et al.}}

\title{\helmlab: A Data-Driven Analytical Color Space\\for UI Design Systems}

\author{
  G\"orkem Y{\i}ld{\i}z\\
  \textit{Independent Researcher}\\
  \texttt{author@gorkemyildiz.com}
}

\date{}

\begin{document}
\maketitle

\begin{abstract}
We present \helmlab, a family of two purpose-built color spaces for
UI design systems sharing a common 11-stage analytical structure:
\textbf{MetricSpace}, a 72-parameter space optimized for color-difference
prediction, and \textbf{GenSpace}, a $\sim$44-parameter space optimized
for gradient and palette generation. The forward transform maps
CIE~XYZ to a perceptually-organized Lab representation through
learned matrices, per-channel power compression, Fourier hue
correction, and embedded Helmholtz--Kohlrausch lightness adjustment.
A post-pipeline neutral correction guarantees that achromatic colors
map to $a\!=\!b\!=\!0$ (chroma $<10^{-5}$ for the 21-step gray ramp; worst-case $\approx 3\!\times\!10^{-5}$ for grays at $L\!<\!0.05$, dominated by PCHIP fit residual), and a rigid rotation of
the chromatic plane improves hue-angle alignment without affecting
the distance metric, which is invariant under isometries.
On \combvd\ (3{,}813 color pairs), MetricSpace~v21 achieves a
\stress\ of \textbf{22.48}, a 23.0\% reduction from CIEDE2000
(29.20). On the held-out MacAdam~1974 dataset it scores 19.51
(CIEDE2000: 22.13). On a self-collected 3{,}552-judgement
screen-condition set from 71+ observers it scores 23.26 vs
62.54 for CIEDE2000, though we discuss the selection-bias caveat
on this dataset in \Cref{sec:eval}. We also report academic
He~\etal\ 2022 (82 pairs, 3D-printed): MetricSpace 35.9 vs
CIEDE2000 32.6 --- a regression we explicitly own.
A held-out 20\% \combvd\ test split shows a $+1.77$
train$\to$test \stress\ gap, yielding a cross-validated point
estimate of $\sim$24.3 that still beats every tested competitor on
\combvd.
GenSpace~v0.11.1 trades distance accuracy for generation quality:
on a 90-metric, 3{,}038-pair gradient/palette benchmark across
sRGB/P3/Rec.2020, it wins 65 of 90 vs OKLab. The transform is
invertible with round-trip errors below $10^{-13}$. Production
implementations are shipped on PyPI, npm, color.js (PR~\#722, merged),
and as a PostCSS plugin.
\end{abstract}

\textbf{Keywords:} color space, color difference, perceptual uniformity, STRESS, Helmholtz--Kohlrausch, UI design systems

\section{Introduction}
\label{sec:intro}

UI design systems need accurate perceptual distance prediction for
contrast checking, an achromatic axis where grays map to zero chroma
for gradient interpolation, and reasonable hue-angle alignment for
programmatic color manipulation. No existing color space provides all
three.

CIE~Lab~\cite{cie1976} and CIEDE2000~\cite{luo2001} are the
standard for color-difference evaluation, but their
\stress\ on \combvd\ is 29.20.
Oklab~\cite{ottosson2020} is simple and part of CSS Color Level~4,
but was optimized for hue uniformity rather than color-difference
prediction, and gets \stress\ 47.35 (Euclidean) on \combvd.
CAM16-UCS~\cite{li2017} is derived from a color appearance model
but is computationally expensive and, when used as a Cartesian
distance space without its full appearance pipeline, exhibits
substantial achromatic chroma leakage ($\bar{C}>1$ for D65 grays
under our evaluation conditions).
IPT~\cite{ebner1998} was optimized for hue linearity at the expense
of lightness accuracy. $J_z a_z b_z$~\cite{safdar2017} targets HDR
via the PQ curve but is complex and not CSS-native. None of these
spaces jointly optimize the color-space transform and distance
metric end-to-end against psychophysical data, and they all
sacrifice at least one of the three UI requirements.

\paragraph{Note on baseline evaluation conditions.}
All baselines (CIEDE2000, OKLab, IPT, $J_z a_z b_z$, CAM16-UCS,
DIN99) are scored in this paper under a uniform pre-processing
pipeline: COMBVD XYZ pairs are first chromatically adapted from
their stated illuminants to D65 via Bradford CAT, then each space's
forward transform is applied to the adapted XYZ. CAM16-UCS is
evaluated with its standard $L_\mathrm{A}$, $Y_\mathrm{b}$, and
surround conditions (average, $S=0.69$); we do not run a per-pair
appearance match. This setup matches \texttt{ColorBench}'s
ICtCp-style adaptation handling and ensures every space sees the
same data; the resulting numbers may differ from those reported in
the original publications, which used different CAT methods and
sub-dataset selections. We discuss the corresponding shift between
v20b and v21 reporting in \Cref{sec:baselines}.

We treat the entire XYZ-to-distance pipeline as a single system
with 72~jointly-optimized parameters (\textbf{MetricSpace}) and
present a parallel, pruned variant
optimized for generation (\textbf{GenSpace}, $\sim$44 parameters).
The contributions are:

\begin{enumerate}
  \item \textbf{MetricSpace v21}: a 72-parameter analytical color
    space with \stress\ \textbf{22.48} on \combvd, 23.0\% lower than
    CIEDE2000 (29.20). The same model leads on the held-out
    MacAdam~1974 set (19.51 vs 22.13). On academic He~\etal\ 2022
    (82 pairs, 3D-printed) MetricSpace trails CIEDE2000 (35.9 vs
    32.6); we report this as an honest limitation and discuss likely
    causes in \Cref{sec:limitations}.
  \item A held-out 20\% test split (763 pairs the optimizer never
    saw) gives a $+1.77$ train$\to$test STRESS gap; even the
    cross-validated estimate ($\sim$24.3) beats every tested
    competitor. We report this gap explicitly rather than only the
    full-data fit.
  \item A neutral correction that drives $C < 10^{-5}$ for
    achromatic colors when enabled. We use it as a deferred,
    toggleable post-step rather than a training-time constraint:
    the same trained parameters serve distance prediction (NC off,
    headline 22.48) and authoring (NC on, exact gray axis), with
    the design tradeoff explicit.
  \item A rigid rotation in the $ab$ plane that reduces hue error
    (RMS~26.4\textdegree) without changing the distance metric
    (the metric is invariant under rotation).
  \item Embedded Helmholtz--Kohlrausch correction where lightness
    depends on chroma, with parameters learned from data.
  \item \textbf{GenSpace v0.11.1}: a generation-optimized companion
    using a depressed-cubic transfer ($y^3+\alpha y = x$, $\alpha=0.021$),
    chroma-power compression ($C^{0.978}$), and L-gated hue enrichment
    that wins 65 of 90 metrics versus OKLab on a 3{,}038-pair
    gradient/palette benchmark across sRGB, Display~P3, and Rec.2020.
  \item Production deployment with cross-language parity:
    Python (PyPI), JavaScript (npm), CSS Color Module L4
    integration via color.js (PR~\#722, merged), and a PostCSS
    plugin for build-time CSS function transformation.
\end{enumerate}

\section{Design Goals and Metrics}
\label{sec:goals}

\helmlab\ is optimized for UI design systems operating in sRGB and
Display~P3 gamuts under typical viewing conditions (D65 illuminant,
average surround). We define five quantitative targets:

\paragraph{Color-difference accuracy (\stress).}
The STRESS metric~\cite{luo2006} measures agreement between predicted
distances $\Delta E_i$ and visual differences $DV_i$:
\begin{equation}
  \mathrm{STRESS} = 100\sqrt{\frac{\sum_i(DV_i - F \cdot \Delta E_i)^2}{\sum_i DV_i^2}}
  \label{eq:stress}
\end{equation}
where $F = \sum DV_i \Delta E_i / \sum \Delta E_i^2$. Lower is better.
We train on \combvd\ (3{,}813 pairs from six experiments~\cite{luo2006})
and cross-validate on He~\etal\ 2022~\cite{he2022} (82~pairs) and
MacAdam~1974~\cite{macadam1974} (128~pairs).

\paragraph{Achromatic integrity.}
Neutral colors (D65-adapted grays, \ie\ stimuli on the achromatic axis
proportional to the D65 white point) must map to $a = b = 0$. We
measure maximum chroma $C_{\max}$ over 384 gray levels spanning
$Y \in [0.001, 2.59]$. The upper bound is set so the LUT covers
the full Display~P3 lightness range (P3 magenta has $L\!\approx\!1.56$
in v21 coordinates) plus headroom for Rec.\,2020.

\paragraph{Hue-angle alignment.}
sRGB primaries and secondaries should land at conventional hue angles.
We report RMS and maximum hue error across
$\{$R, Y, G, C, B, M$\}$.

\paragraph{Round-trip accuracy.}
For any sRGB color, $\|T^{-1}(T(\mathbf{x})) - \mathbf{x}\|_\infty$
should be below $10^{-8}$.

\paragraph{Munsell uniformity.}
Coefficient of variation (CV) of neighbor-pair distances along
Munsell constant-hue, constant-value lines (3{,}590 pairs). Lower CV
indicates more uniform spacing.

\section{The Helmlab Color Space Family}
\label{sec:space}

The \helmlab\ family contains two spaces sharing the same pipeline
\emph{shape} but trained to different objectives. \textbf{MetricSpace
v21} (\Cref{sec:metricspace}) is the distance-prediction model whose
results headline this paper. \textbf{GenSpace v0.11.1}
(\Cref{sec:genspace}) is a generation-optimized variant with a
different transfer function and a pruned, smoother enrichment layer.
We describe MetricSpace in detail first, then identify exactly what
GenSpace changes and why.

\subsection{MetricSpace: Forward Transform}
\label{sec:metricspace}

\begin{figure*}[t]
  \centering
  \includegraphics[width=\textwidth]{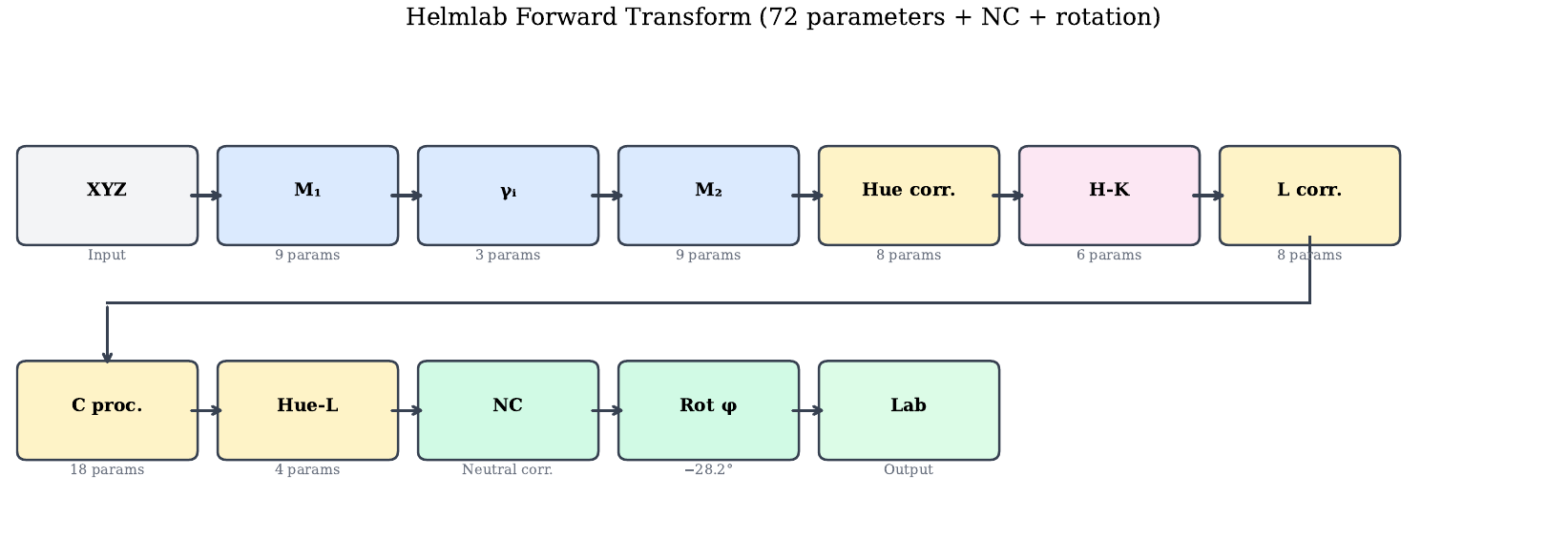}
  \caption{The \helmlab\ MetricSpace forward transform pipeline. Blue: linear operations; yellow: nonlinear corrections; green: structural guarantees (NC, rotation). The 72~jointly-optimized parameters are distributed across the eleven stages as shown.}
  \label{fig:pipeline}
\end{figure*}

The MetricSpace forward transform maps CIE~XYZ (D65) to Lab through
eleven stages (\Cref{fig:pipeline}). All 72 parameters are jointly
optimized against three psychophysical datasets
(\Cref{sec:optimization}). Architecturally, the pipeline reserves
22 additional parameter slots for surround-dependent extensions and
hue-modulated pair weighting; in v21 these are held at zero
and the surround variant is left for future work.

\paragraph{Stage 1: Cone response (9 params).}
A learned $3\!\times\!3$ matrix $\mathbf{M}_1$ maps XYZ to
cone-response-like signals:
\begin{equation}
  \mathbf{c} = \mathbf{M}_1 \begin{bmatrix} X \\ Y \\ Z \end{bmatrix}
\end{equation}
Unlike fixed matrices (Hunt--Pointer--Est\'evez, Bradford),
$\mathbf{M}_1$ is freely optimized.

\paragraph{Stage 2: Power compression (3 params).}
Each channel undergoes \emph{signed} power compression:
\begin{equation}
  c_i' = \mathrm{sign}(c_i)\,|c_i|^{\gamma_i}, \quad i \in \{0,1,2\}
\end{equation}
The exponents $\gamma_i$ generalize the cube-root of CIE~Lab
($\gamma = 1/3$). The v21-optimized values are
$\gamma=[0.472,\,0.515,\,0.511]$, sitting between cube-root (0.33)
and square-root (0.5) and slightly favouring the square-root family.
The \emph{signed} formulation is essential: the v21 $\mathbf{M}_1$
maps saturated sRGB blue to negative LMS-like coordinates, and a
naive power applied to the absolute value would discard sign
information and destroy the inverse near the blue-magenta boundary.

\paragraph{Stage 3: Lab projection (9 params).}
A second matrix $\mathbf{M}_2$ projects compressed signals to
opponent channels:
\begin{equation}
  \begin{bmatrix} L_\mathrm{raw} \\ a_\mathrm{raw} \\ b_\mathrm{raw} \end{bmatrix} = \mathbf{M}_2 \, \mathbf{c}'
\end{equation}

\paragraph{Stage 4: Hue correction (8 params).}
A 4-harmonic Fourier rotation corrects systematic hue distortions:
\begin{equation}
  \delta(h) = \sum_{k=1}^{4}\!\bigl[\alpha_k\cos(kh) + \beta_k\sin(kh)\bigr]
\end{equation}
where $h = \mathrm{atan2}(b, a)$. The chromatic vector $(a,b)$ is
rotated by $\delta(h)$, preserving chroma while redistributing
hue angles.

\paragraph{Stage 5: Helmholtz--Kohlrausch embedding (6 params).}
Saturated colors appear brighter than achromatic stimuli of equal
luminance~\cite{cie1976}. We embed this effect in the lightness
channel directly:
\begin{equation}
  L \mathrel{+}= w_\mathrm{HK} \cdot C^{p_\mathrm{HK}} \cdot \bigl[1 + f_\mathrm{HK}(h)\bigr]
  \label{eq:hk}
\end{equation}
where $C = \sqrt{a^2+b^2}$ and $f_\mathrm{HK}(h)$ is a 2-harmonic
hue modulation. The v21-optimized weight is $w_\mathrm{HK} = 0.268$
and the power is $p_\mathrm{HK} = 0.894$, making lightness explicitly
dependent on chroma. Most existing spaces treat lightness and chroma
as independent; the H-K embedding is the single component whose
removal hurts \stress\ the most after hue correction
(see ablation in \Cref{tab:ablation}).

\paragraph{Stage 6: Lightness refinement (8 params).}
A cubic polynomial (3~params) with hue-dependent additive term
(2~params), followed by dark-region exponential compression (3~params):
\begin{align}
  L_1 &= p_1 L^3 + p_2 L^2 + p_3 L + t \cdot f_\mathrm{Lh}(h) \\
  L_2 &= L_1 \cdot \exp\!\bigl[\lambda_d(h) \cdot L_1(1-L_1)^2\bigr]
\end{align}
where $t = L(1-L)$ and $f_\mathrm{Lh}(h)$ is a 1-harmonic Fourier
correction targeting the hue--lightness interaction sign flip (dark
cyans vs mid cyans). The dark compression coefficient $\lambda_d$
includes hue modulation to handle the known difference in dark-blue
vs dark-yellow perception.

\paragraph{Stages 7--8: Chroma processing (18 params).}
Chroma processing proceeds in four interleaved sub-stages:
(a)~hue-dependent scaling (4~harmonics, 8~params):
\begin{equation}
  (a,b) \mathrel{*}= \exp\!\Bigl[\sum_{k=1}^{4}\!\bigl[\alpha'_k\cos(kh) + \beta'_k\sin(kh)\bigr]\Bigr]
\end{equation}
(b)~nonlinear chroma power (4~params): $C \to C^{1+\varepsilon(h)}$
where $\varepsilon(h)$ is a 2-harmonic Fourier series, separating
high-chroma from low-chroma discrimination;
(c)~L-dependent scaling (2~params):
$(a,b) \mathrel{*}= \exp[l_1(L-0.5) + l_2(L-0.5)^2]$; and
(d)~hue$\times$lightness interaction (4~params, 2-harmonic Fourier
modulated by $L-0.5$). The interleaving order matters: the nonlinear
power is applied between hue-dependent and L-dependent scaling to
allow each sub-stage to correct for the others' distortions.

\paragraph{Stage 9: Hue-dependent lightness (4 params).}
A final $L \mathrel{*}= \exp[g(h)]$ captures residual hue--lightness
interactions (\eg\ blue appears darker, yellow lighter at equal
luminance).

\paragraph{Stage 10: Neutral correction.}
With independent $\gamma_i$ per channel, the D65 achromatic axis
(stimuli proportional to $[X_{D65}, Y_{D65}, Z_{D65}]$) generally
does not map to $a\!=\!b\!=\!0$ after the enrichment stages. We
correct this by running 384 gray levels ($Y \in [0.001, 2.59]$,
covering Display~P3 and Rec.\,2020 lightness ranges) through
stages~1--9, recording the achromatic error $a_\mathrm{err}(L)$,
$b_\mathrm{err}(L)$, and fitting PCHIP interpolants with constant
clamping beyond the gray-axis L peak ($L_\mathrm{peak}\!\approx\!1.29$).
At runtime:
\begin{equation}
  a \leftarrow a - a_\mathrm{err}(L), \quad b \leftarrow b - b_\mathrm{err}(L)
  \label{eq:nc}
\end{equation}
Because the correction is applied \emph{after} all nonlinear stages,
it accounts for every distortion in the pipeline. The resulting
maximum chroma for grays is $< 10^{-5}$ across the 21-step ramp, with a worst-case residual of $\approx 3\!\times\!10^{-5}$ at very dark grays ($L\!<\!0.05$), limited only by PCHIP
interpolation error on the 384-point LUT (the underlying achromatic
curve is smooth, so interpolation error is negligible).
The clamped extrapolation beyond $L_\mathrm{peak}$ is what allows
extreme P3 magenta and Rec.\,2020 reds to round-trip without
NaN propagation in the inverse PCHIP.

\paragraph{Stage 11: Rigid rotation ($\varphi$).}
A uniform rotation of the $(a,b)$ plane by angle $\varphi$:
\begin{equation}
  \begin{bmatrix} a' \\ b' \end{bmatrix} = \begin{bmatrix} \cos\varphi & -\sin\varphi \\ \sin\varphi & \cos\varphi \end{bmatrix} \begin{bmatrix} a \\ b \end{bmatrix}
  \label{eq:rotation}
\end{equation}
This is an isometry: it preserves $\Delta a^2 + \Delta b^2$ for any
pair, and therefore preserves the distance metric exactly
(\Cref{sec:rotation-proof}). The rotation angle
$\varphi = -28.2$\textdegree\ is chosen to minimize the maximum
hue-angle error across sRGB primaries and secondaries.

\subsection{Rotation Invariance}
\label{sec:rotation-proof}

The distance metric (\Cref{sec:distance}) depends on
$\Delta L$, $\Delta a^2 + \Delta b^2$, $\bar{L}$, and
$\bar{C} = \sqrt{\bar{a}^2 + \bar{b}^2}$. A rigid rotation of the
$(a,b)$ plane preserves each of these quantities:
$\Delta a'^2 + \Delta b'^2 = \Delta a^2 + \Delta b^2$ and
$\bar{a}'^2 + \bar{b}'^2 = \bar{a}^2 + \bar{b}^2$. Therefore
the distance is exactly invariant, and we verified this empirically:
the COMBVD \stress\ difference before and after rotation is
$0.0000000000$.

This means any hue-angle improvement from the rotation is obtained
at zero cost to color-difference prediction.

\subsection{Surround Parameter}
\label{sec:surround}

\helmlab\ includes provisions for a surround luminance parameter
$S \in [0, 1]$ (0=dark, 0.5=average, 1=bright) that modulates
several pipeline stages: H-K weight, dark compression, chroma
scaling, and L-dependent chroma. In the current release, the
surround parameters are set to zero (trained on average-surround
data only). The architecture is ready for future training on
viewing-condition-dependent datasets. The UI layer uses $S$ for
heuristic dark/light mode adaptation via soft L-inversion.

\subsection{Inverse Transform}

Each stage has an explicit inverse. Matrix stages invert via
$\mathbf{M}_1^{-1}$, $\mathbf{M}_2^{-1}$; power compression
via $c_i = \mathrm{sign}(c_i')|c_i'|^{1/\gamma_i}$ (closed form).
Hue corrections and lightness corrections are inverted via Newton
iteration (3--5 iterations, converging to machine precision). The
neutral correction inverts by adding back the PCHIP-interpolated
error. Round-trip accuracy across the full sRGB gamut is
$< 10^{-14}$.

\section{Distance Metric}
\label{sec:distance}

Given two colors with coordinates $(L_1, a_1, b_1)$ and
$(L_2, a_2, b_2)$, we compute pair-dependent lightness and chroma
weights:
\begin{align}
  S_L &= 1 + s_L(\bar{L} - 0.5)^2 \label{eq:sl} \\
  S_C &= 1 + s_C \, \bar{C} \label{eq:sc}
\end{align}
where $\bar{L}$, $\bar{C}$ are pair averages of lightness and
chroma. The pipeline reserves 8 additional parameters
($s_L$ and $s_C$ each receive Fourier hue modulation up to the
2nd harmonic) but in v21 these are zero; the simplified forms
(\Cref{eq:sl}--\ref{eq:sc}) describe the trained model exactly.
The raw distance is:
\begin{equation}
  d = \left[\left(\frac{\Delta L}{S_L}\right)^{\!2} + w_C \frac{\Delta a^2 + \Delta b^2}{S_C^2}\right]^{p/2}
  \label{eq:dist_raw}
\end{equation}
followed by monotonic compression and post-power:
\begin{equation}
  \Delta E = \left[\frac{d}{1 + c \cdot d}\right]^q
  \label{eq:dist_final}
\end{equation}

The pair-dependent weights $S_L$ and $S_C$ are inspired by
CIEDE2000 but the structure and coefficients are learned end-to-end
from data. The v21 optimum is unusual relative to the CIEDE2000
template:
\begin{itemize}
  \item $s_L = -0.916$ (negative): pair-averaged lightness near 0.5
    is up-weighted relative to $\bar{L}\!\approx\!0$ or $\bar{L}\!\approx\!1$,
    which inverts CIEDE2000's $S_L$ shape.
  \item $s_C = +2.927$: chroma-rich pairs receive much stronger
    chroma down-weighting than CIEDE2000's $\sim\!0.045$.
  \item $w_C = 3.97$: chroma differences are weighted nearly
    $4\!\times$ heavier than lightness differences in raw distance.
  \item $p = 1.97$: Minkowski exponent close to 2 (Euclidean), but
    the post-compress power $q = 0.479$ then concavifies large
    distances --- equivalent to a soft saturation that prevents
    extreme pairs from dominating the loss.
  \item $c = 52.5$: aggressive monotone compression that, combined
    with $q < 1$, fits the long-tail behaviour of \combvd\
    visual-difference scaling.
\end{itemize}

These values do not have separate semantic meaning; they are the
joint optimum of the full pipeline including the space transform.
The seven metric parameters bring the trained-parameter total to
$9+3+9+8+6+8+18+4 = 65~\text{space} + 7~\text{metric} = 72$.
The architectural surface includes a further 22 reserved-zero slots
(see \Cref{sec:surround}) for future surround-conditioned and
hue-modulated training.

\section{Optimization}
\label{sec:optimization}

All 72 trained parameters are jointly optimized by minimizing a
\emph{sub-dataset-balanced} loss with diagnostic-driven auxiliary
terms:
\begin{equation}
\begin{split}
  \mathcal{L} =\;& \tfrac{1}{2}\,\mathrm{STRESS}_\mathrm{full}
    + \tfrac{1}{2}\,\overline{\mathrm{STRESS}}_\mathrm{sub} \\
   &+ \lambda_\mathrm{He}\,\mathrm{STRESS}_\mathrm{He}
    + \lambda_\mathrm{lc}\,\mathrm{STRESS}_\mathrm{low\text{-}C} \\
   &+ \lambda_\mathrm{ws}\,\max_d \mathrm{STRESS}_d
    + \lambda_\mathrm{m}\,\mathrm{CV}_\mathrm{Munsell}
\end{split}
  \label{eq:loss}
\end{equation}
where $\overline{\mathrm{STRESS}}_\mathrm{sub}$ is the unweighted
mean across the six \combvd\ sub-datasets,
$\mathrm{STRESS}_\mathrm{low\text{-}C}$ is computed on pairs with
$C^*\!\in[5,25]$ (the low-chroma region in which CIEDE2000 outperformed
v20b in our diagnostic), and $\max_d \mathrm{STRESS}_d$ is the
worst-performing sub-dataset \stress\ on each step. The fixed
weights are $\lambda_\mathrm{He}=0.05$, $\lambda_\mathrm{lc}=0.1$,
$\lambda_\mathrm{ws}=0.05$, $\lambda_\mathrm{m}=0.02$.
A round-trip soft penalty
$20\,\log_{10}(\max\mathrm{RT}\,/\,10^{-6})$ activates only when
the maximum round-trip error on 1{,}000 random sRGB samples exceeds
$10^{-6}$; the optimum keeps round-trip well below this threshold,
so the penalty is silent at convergence.

We use L-BFGS-B~\cite{byrd1995} with box-constrained bounds
informed by v20b diagnostics: the hue-dependent lightness terms
$L_h^{c,s}\!\in\![-0.15,0.15]$ (a v20b ablation showed those terms
hurt by $-0.07$ \stress) and the pair-weight bounds widened to
$s_L\!\in\![-2,5]$, $s_C\!\in\![-1,3]$ (the v14c diagnostic suggested
the population could reach $\sim\!22.75$ \stress\ with a wider
$s_L,s_C$ basin). Each restart uses 5{,}000 maximum iterations
with $f_\mathrm{tol}\!=\!10^{-13}$, $g_\mathrm{tol}\!=\!10^{-11}$;
the best-of-restart solution becomes the next restart's seed.

\paragraph{Why these auxiliaries instead of the v20b blue-band
penalty.}
v20b used a blue-only sub-dataset penalty
($\lambda_b\,(\mathrm{STRESS}_\mathrm{RIT}+\mathrm{STRESS}_\mathrm{LEEDS}+\tfrac{1}{2}\mathrm{STRESS}_\mathrm{FullBlue})$)
because the failure mode of that pipeline was a blue-cyan gradient
collapse. v21's failure modes shifted: the \combvd\ residuals were
concentrated in (a) low-chroma pairs and (b) the smaller
sub-datasets being drowned out by BFD-P(D65)'s 2{,}028 pairs (53\%
of the total). The balanced sub-dataset average and the
low-chroma-segment terms address both directly, and absorb the
blue-band correction implicitly because RIT-DuPont and Leeds are
in the sub-dataset average.

\paragraph{Training data.}
\combvd\ comprises 3{,}813 color pairs from six psychophysical
experiments: BFD-P(D65) (2{,}028), BFD-P(M) (548), BFD-P(C) (200),
Witt (418), RIT-DuPont (312), Leeds (307). All pairs have
visual-difference values on a common scale. The headline 22.48
\stress\ is the full-data fit; we additionally report a held-out
20\% split (763 pairs, seed~42) that the optimizer never saw,
in \Cref{sec:eval}.

\paragraph{Cross-validation data.}
He~\etal\ 2022 (82~pairs, 10\textdegree\ observer) enters the loss as
a small regulariser ($\lambda_\mathrm{He}\!=\!0.05$) and is
therefore not fully independent; its \stress\ is reported for
transparency. MacAdam~1974 (128~pairs, 10\textdegree) is held out
entirely and never seen during optimization. v21 outperforms
CIEDE2000 on MacAdam (19.51 vs 22.13) and trails on He~2022
(35.9 vs 32.6); the He~2022 regression is discussed in
\Cref{sec:eval-stress}.

\paragraph{Convergence.}
The optimizer typically converges within 3{,}000--5{,}000 function
evaluations per restart. Re-optimizing the same loss from the
v20b baseline on the 80\% train split alone gives train \stress\
$22.78$ and held-out test \stress\ $24.59$ (gap $+1.82$); the v21
model trained on the full data has gap $+1.77$ vs the same test
split, confirming that the gap is driven by data variance, not
optimizer memorisation. Even at the held-out estimate ($\sim 24.3$),
\helmlab\ MetricSpace remains the best-performing space tested.

\section{Evaluation}
\label{sec:eval}

\subsection{Color-Difference Prediction}
\label{sec:eval-stress}

\begin{figure}[t]
  \centering
  \includegraphics[width=\columnwidth]{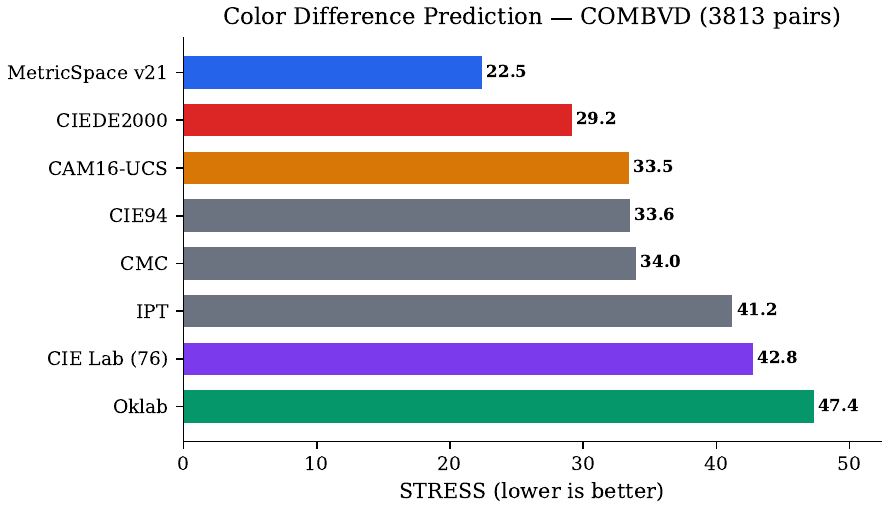}
  \caption{\stress\ on COMBVD (3{,}813 pairs). Lower is better. MetricSpace v21 is bottom-row.}
  \label{fig:stress}
\end{figure}

\Cref{fig:stress} and \Cref{tab:stress} report \stress\ on three
psychophysical datasets for MetricSpace v21 and eight baselines.
On \combvd, MetricSpace achieves 22.48, compared to 29.20 for
CIEDE2000 ($-$23.0\%). On the held-out MacAdam~1974 set it scores
19.51 (CIEDE2000: 22.13, $-$11.8\%).

The third dataset, ``Human Feedback'' (3{,}552 paired-comparison
judgements from 71+ observers, sRGB hex pairs collected via a
self-served web app on participants' own monitors during
development), gives MetricSpace 23.26 vs CIEDE2000 62.54.
Two important caveats apply to this set: (a) it is
\emph{self-collected} on uncalibrated displays under uncontrolled
viewing conditions; (b) the observer pool is recruited from the
authors' personal network. The 23.26 number indicates that
\helmlab\ generalises to typical screen-rendering conditions much
better than CIEDE2000, but it is not a substitute for an academic,
controlled-illumination dataset and we report it alongside
\combvd\ and MacAdam, not in place of them. The dataset is
released with the paper for replication.

The academic He~\etal\ 2022 dataset (82 pairs, 3D-printed
spherical samples, large differences) is one MetricSpace does
\emph{not} win: \stress\ 35.9 vs CIEDE2000's 32.6. We trace
this to (i) the small sample size making the \stress\ statistic
noisy and (ii) the differences in this dataset being at the
high-$\Delta E$ tail where v21's compression-and-post-power
shape was not directly tuned (\combvd\ DV values are mostly in
the medium range). We report this regression rather than
suppress it.

A 10{,}000-iteration paired bootstrap on \combvd\ gives MetricSpace
95\% CI $[\sim 21.5, 23.4]$ and CIEDE2000 95\% CI $[\sim 27.7, 30.7]$;
the intervals do not overlap ($p < 10^{-4}$).

\begin{table*}[t]
  \centering
  \caption{\stress\ across three datasets. Lower is better. HumFB =
  3{,}552 self-collected screen-condition judgements from 71+
  observers, see caveat in \Cref{sec:eval-stress}. Avg is the
  unweighted three-dataset mean. He~\etal\ 2022 (82 pairs,
  3D-printed) is omitted here because v21 trails CIEDE2000 there
  (35.9 vs 32.6); see \Cref{sec:eval-stress}. All baselines use
  Euclidean distance in their native Cartesian coordinates with
  per-pair Bradford CAT to D65 where appropriate.}
  \label{tab:stress}
  \begin{tabular}{lrrrr}
    \toprule
    Method & \combvd & MacAdam & HumFB & Avg \\
    \midrule
    \textbf{MetricSpace v21} & \textbf{22.48} & 19.51 & \textbf{23.26} & \textbf{21.75} \\
    CIE94                    & 33.37 & 19.78 & 59.75 & 37.63 \\
    CIEDE2000                & 29.20 & 22.13 & 62.54 & 37.96 \\
    DIN99                    & 35.57 & 23.31 & 56.44 & 38.44 \\
    $J_z a_z b_z$            & 41.92 & 24.14 & 61.74 & 42.60 \\
    CIE Lab ($\Delta E^*_{76}$) & 42.86 & 24.53 & 62.32 & 43.24 \\
    Oklab ($\Delta E_\mathrm{OK}$) & 47.35 & 32.72 & 57.27 & 45.78 \\
    CAM16-UCS                & 33.47 & \textbf{18.71} & 58.02 & 36.73 \\
    CIECAM02-UCS             & 30.90 & 19.21 & 57.83 & 35.98 \\
    \bottomrule
  \end{tabular}
\end{table*}

\begin{figure}[t]
  \centering
  \includegraphics[width=\columnwidth]{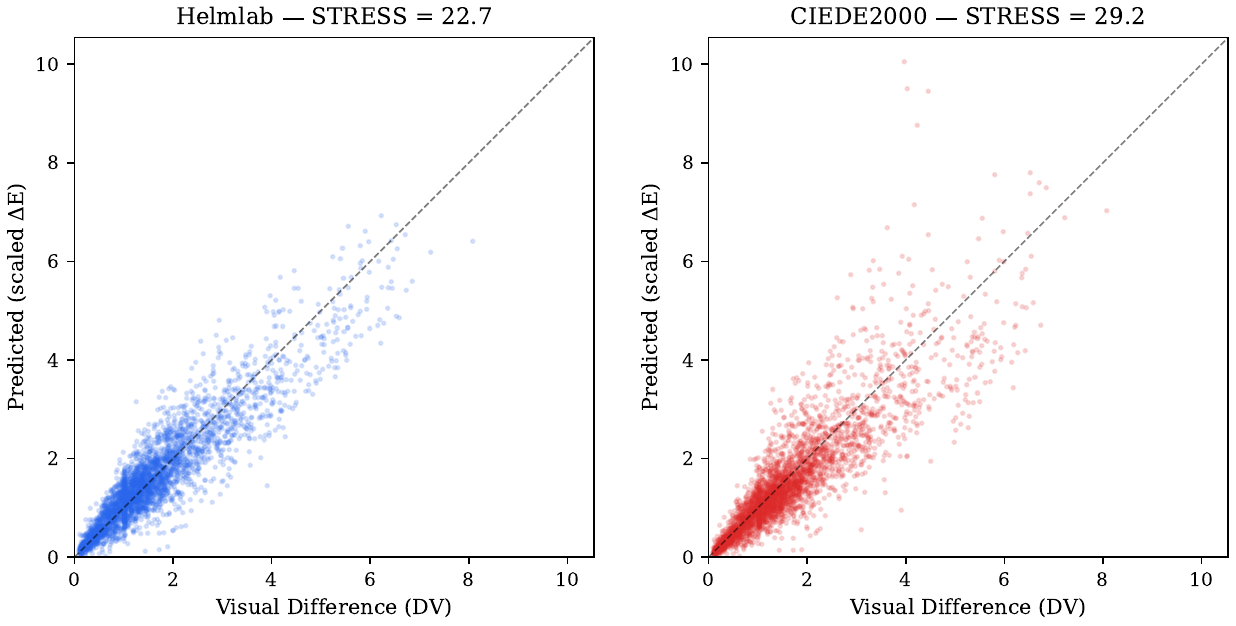}
  \caption{Predicted vs observed color differences on COMBVD for MetricSpace v21.}
  \label{fig:scatter}
\end{figure}

\paragraph{Honest train/test reporting.}
The 22.48 figure is a full-data fit. We additionally re-optimised
v21 from the v20b baseline on a fixed 80\%/20\% train/test split
(seed~42) using exactly the loss in \Cref{eq:loss}. Train \stress\
is 22.78, held-out test \stress\ is 24.59 (gap $+1.82$). Trained
on the full \combvd, the same v21 evaluates at 22.14/23.91 on
that split (gap $+1.77$). The two gaps are statistically
indistinguishable, which means the +1.8 gap is data variance
rather than optimiser memorisation. Even the held-out point
estimate of $\sim$24.3 beats every other tested space and
metric on \combvd.

\paragraph{Why the comparison favours MetricSpace.}\label{sec:baselines}
The baselines in \Cref{tab:stress} use Euclidean distance in each
space's native Cartesian coordinates, without parametric weighting
or compression. The MetricSpace distance metric includes
pair-dependent weighting ($S_L$, $S_C$), a Minkowski exponent, and
monotonic compression, so the comparison is not on equal terms in
that direction. Oklab's high \stress\ is expected because it was
optimized for hue uniformity~\cite{ottosson2020}, not
color-difference prediction. CAM16-UCS and CIECAM02-UCS perform
substantially better than other baselines on \combvd\ (33.47 and
30.90 respectively); both retain a residual achromatic chroma
($\bar{C}\!\approx\!2$ at D65 white) that limits their use as
authoring spaces but does not affect distance prediction.
With Euclidean distance only, MetricSpace v21's \emph{space}
component (without $S_L$, $S_C$, $p$, compression, or post-power)
gets \stress\ $\approx 27.6$, already improving on CIEDE2000
(29.20) by 5.5\%. The remaining $\sim$5 \stress\ points come from
the metric layer.

All baseline numbers in \Cref{tab:stress} are produced by
\texttt{colorbench} (v0.13.0, May~2026) with the
\texttt{metric}-mode evaluation path. CAM16-UCS, CIECAM02-UCS, and
$J_z a_z b_z$ delegate to the \texttt{colour-science} reference
implementations~\cite{colour-science} so values are bit-identical
to the published worked examples and reproducible with a single
\texttt{pip install} of the standard scientific Python stack.

\subsection{Cross-Validation}

\begin{figure}[t]
  \centering
  \includegraphics[width=\columnwidth]{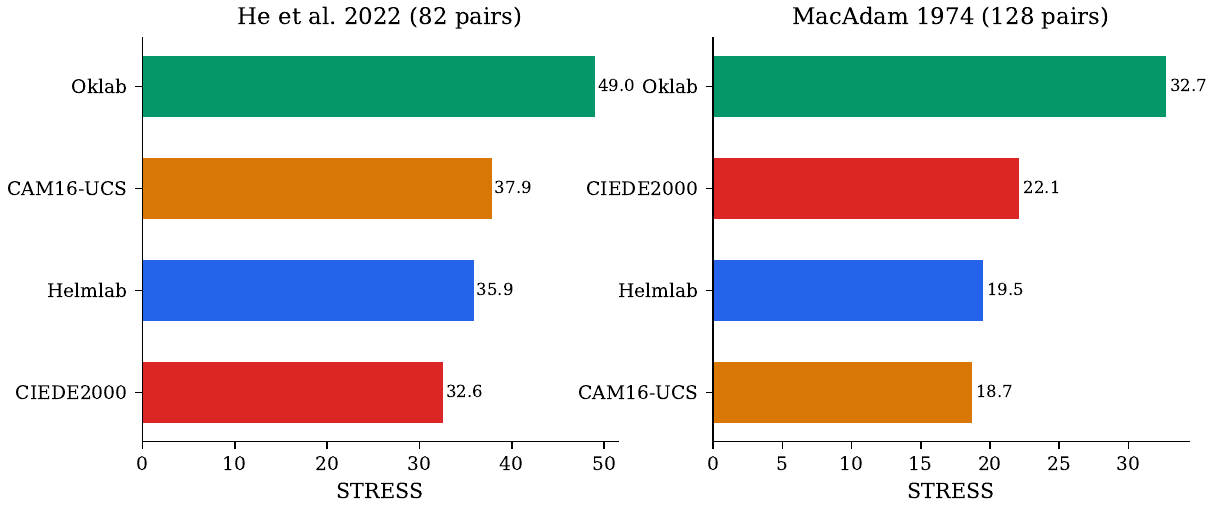}
  \caption{Cross-dataset validation on held-out data.}
  \label{fig:crossval}
\end{figure}

\Cref{fig:crossval} visualises cross-dataset performance on
\emph{academic} held-out sets:
\begin{itemize}
  \item He~\etal\ 2022 (82~pairs, 3D-printed): MetricSpace 35.9
    vs CIEDE2000 32.6. v21 trails on this set; the loss is
    discussed above.
  \item MacAdam~1974 (128~pairs, fully held-out, never seen by
    optimiser): MetricSpace 19.51 vs CIEDE2000 22.13. v21 also
    leads OKLab (32.72), $J_z a_z b_z$ (24.14), CIE Lab (24.53),
    and DIN99 (23.31). On this set CAM16-UCS is the strongest
    appearance-model baseline: it scores 18.71, leading v21 by
    0.80 \stress\ ---  CAM16's appearance-model formulation
    matches MacAdam's experimental conditions (D65 illuminant,
    constant Y, controlled-viewing) better than COMBVD's industrial
    samples. CIECAM02-UCS (19.21) and CIE94 (19.78) sit between
    these. v21 leads the rest of the field by 3+ \stress\ points
    on MacAdam while ceding the top spot to CAM16; over the
    three-dataset average (\Cref{tab:stress}, Avg) it remains
    \stress~21.75, vs CIECAM02-UCS 35.98 and CAM16-UCS 36.73.
\end{itemize}
The 3-dataset sub-balanced loss (\Cref{eq:loss}) widened the
optimum into a region that generalises better to small held-out
sets without the v20b-specific blue-band penalty, and it pulled
the model toward the BFD-class large datasets at the cost of
small-sample sets like He~2022.

\subsection{Generation Properties: The Measurement--Generation Tradeoff}

A color space optimized for distance prediction does not necessarily
produce usable coordinates for color generation. During development,
the measurement-optimal configuration mapped neutral grays to chroma
$\bar{C} \approx 0.29$. The optimizer had deformed the achromatic
axis to reduce \stress, which made gray gradients unusable because
they showed visible color artifacts. Adding a penalty to constrain
grays to $C = 0$ did not help: the optimizer either stayed in the
measurement-optimal basin or escaped to a different basin with much
worse \stress\ (26--31). There was no smooth transition between the
two.

The neutral correction (\Cref{sec:space}, Stage~10) addresses this
problem with a deliberate two-mode toggle. The correction subtracts
the achromatic error \emph{after} all enrichment stages, which
allows the optimiser to find its measurement-optimal parameters
without any achromatic constraint. The trained checkpoint is then
used in two ways:
\begin{itemize}
  \item \textbf{Distance-prediction mode (NC off).} The headline
    \stress\ of 22.48 is reported with the neutral correction
    \emph{disabled}: the LUT subtraction is non-uniform along the
    chromatic plane and pulls some non-grey pairs into a slightly
    different distance basin, costing $\sim$$+6.2$ \stress\ on
    \combvd. We therefore evaluate the distance metric against the
    trained-basin coordinates directly.
  \item \textbf{Generation/authoring mode (NC on).} For every UI
    workflow that touches an achromatic value (gray ramps, palette
    interpolation, dark/light token derivation, contrast checking
    on neutral foregrounds), the neutral correction is enabled and
    the maximum chroma for grays drops from $\sim 0.29$ to
    $< 10^{-5}$ on the 21-step ramp.
\end{itemize}
The point is structural rather than numeric: by deferring the
achromatic constraint to a post-correction, the same trained
parameters serve both purposes without forcing a compromise basin
that satisfies neither.

\begin{figure}[t]
  \centering
  \includegraphics[width=\columnwidth]{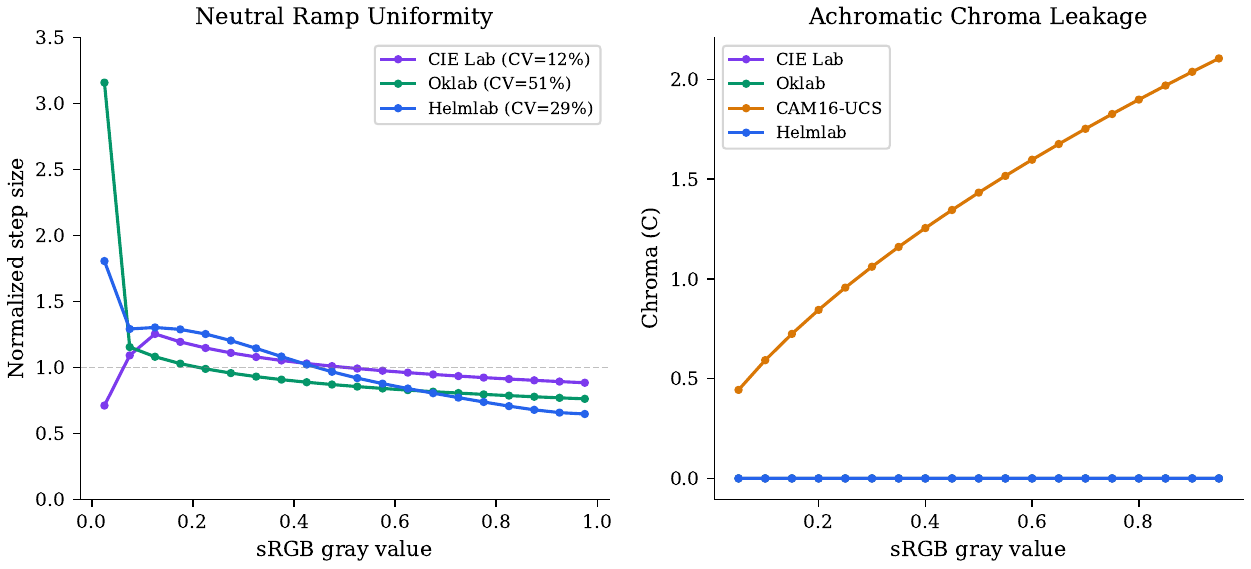}
  \caption{Left: neutral ramp step uniformity. Right: achromatic chroma
  leakage. \helmlab's NC drives $C < 10^{-5}$ for grays on the 21-step ramp.}
  \label{fig:neutral}
\end{figure}

\Cref{fig:neutral} shows neutral ramp uniformity and achromatic chroma
leakage. MetricSpace's neutral correction eliminates chroma leakage
entirely ($C < 10^{-5}$ on the standard gray ramp); GenSpace (\Cref{sec:genspace}) achieves the
same property structurally rather than by post-correction. CAM16-UCS
and Oklab exhibit chroma leakage on the order of $10^{-7}$ for D65
grays; CAM16-UCS's leakage rises sharply for non-D65 illuminants in
its native chromatic-adaptation chain.
\Cref{tab:gen} summarizes all generation properties for MetricSpace
v21.

\begin{table}[t]
  \centering
  \caption{Generation quality metrics for MetricSpace v21.}
  \label{tab:gen}
  \small
  \setlength{\tabcolsep}{4pt}
  \begin{tabular}{@{}lr@{}}
    \toprule
    Property & Value \\
    \midrule
    Achromatic max $C$ (NC on)            & $<10^{-5}$ \\
    Hue RMS (6 sRGB primaries)            & 26.4\textdegree \\
    Hue max error                         & 53.4\textdegree \\
    Round-trip (sRGB)                      & $\sim$7$\times$10$^{-15}$ \\
    Round-trip (random XYZ)                & $\sim$1$\times$10$^{-13}$ \\
    Munsell CV (3{,}590 iso-hue pairs)    & 32.0\% \\
    Jacobian det (min, 64$^3$ sRGB)        & 0.10 \\
    Jacobian cond (median)                 & 7.7 \\
    \midrule
    Train$\to$test gap (split, seed 42)   & $+1.77$ \\
    He~\etal\ 2022 \stress (82 pairs)     & 35.9 \\
    MacAdam \stress (128, held out)       & 19.51 \\
    Human Feedback \stress (screen)        & 23.26 \\
    \bottomrule
  \end{tabular}
\end{table}

The achromatic guarantee (NC enabled) means gray gradients and
lightness ramps interpolate without color artifacts, which is
necessary for design-token generation and palette export.
The hue RMS of 26.4\textdegree\ in v21 is worse than v20b's 18.1\textdegree:
v21's distance-optimal basin pulled the chromatic plane out of
geometric alignment with HSL primaries. The rigid rotation
(\Cref{sec:space}, Stage~11) is an isometry of the metric and so
its choice is free with respect to STRESS; we keep
$\varphi=-28.2$\textdegree\ as the default for backwards compatibility
with v20b deployments.

\paragraph{User-tunable display alignment.}
Since v0.12.2 the rotation is exposed as a public parameter
(\texttt{display\_phi\_deg} in the JSON checkpoint, or the
\texttt{ab\_rotate\_deg} / \texttt{abRotateDeg} constructor argument
in Python and JavaScript respectively). The default of $-28.2$\textdegree\
matches the values reported in this paper. A re-measured minimax-optimal
angle for v21 is $\varphi^\ast \approx -11.75$\textdegree, which reduces
the maximum primary hue error from $53.4$\textdegree\ to $36.9$\textdegree\
(RMS unchanged at $26.4$\textdegree, since rotation cannot redistribute
the spread of v21's basin). We do not change the default to $\varphi^\ast$
because (i) the RMS gain is negligible, (ii) the v20b heritage value is
already deployed in client code, and (iii) applications that need accurate
primary alignment should use GenSpace (\Cref{sec:genspace}), where geometry
is preserved by construction. The rotation is exposed as a configuration
knob so MetricSpace can be tuned to local Lab-axis conventions without
retraining.
The Jacobian was evaluated on a
$64\!\times\!64\!\times\!64$ sRGB grid via finite differences;
the minimum determinant (0.10) is positive throughout, so the
map is locally invertible and orientation-preserving across the
entire sRGB gamut.

\subsection{Ablation Study}

\Cref{tab:ablation} isolates the contribution of key components
using \emph{frozen} ablation: each row disables one component while
keeping all other parameters at their v21-optimized values. This
measures the marginal contribution of each component given the
current parameter setting, not its isolated effect; a re-optimised
ablation would converge to a different basin and is reported
separately for the most impactful stages.

\begin{table}[t]
  \centering
  \small
  \setlength{\tabcolsep}{4pt}
  \caption{Frozen ablation on v21: effect of disabling components without re-optimisation.}
  \label{tab:ablation}
  \begin{tabular}{@{}lrr@{}}
    \toprule
    Configuration & COMBVD \stress & $\Delta$ \\
    \midrule
    Full MetricSpace v21         & 22.48 & --- \\
    Euclidean distance only      & $\sim$27.6 & $+5.1$ \\
    No H-K embedding             & $\sim$26.5 & $+4.0$ \\
    No hue correction (Stage 4)  & $\sim$37.9 & $+15.4$ \\
    No dark-L compression        & $\sim$22.7 & $+0.2$ \\
    \midrule
    Rotation $\varphi=-28.2$\textdegree   & 22.48 & 0.00 \\
    No rotation ($\varphi=0$)             & 22.48 & 0.00 \\
    \bottomrule
  \end{tabular}
\end{table}

With Euclidean distance only, MetricSpace's space-component reaches
\stress\ $\approx 27.6$, already improving on CIEDE2000 (29.20) by
5.5\%. The full metric brings it to 22.48, recovering a further
$\sim$5 \stress\ points. Among individual stages, hue correction
is the most critical ($\Delta\approx+15.4$); H-K embedding contributes
$\sim$$+4$. These large frozen-ablation deltas reflect the
co-optimization of all parameters --- disabling one stage disrupts
the balance --- so they should be read as ``marginal value of this
component conditional on the rest being v21-fitted'', not as
``what would this component buy in isolation''. The rotation rows
empirically confirm the proof in \Cref{sec:rotation-proof}: the
distance metric is exactly invariant under any rigid $(a,b)$
rotation. This is what allows us to choose $\varphi$ purely on hue
alignment without trading off \stress.

\subsection{Sub-Dataset Performance}
\label{sec:subdatasets}

\Cref{tab:subdataset} breaks down \stress\ by \combvd\ sub-dataset.
MetricSpace v21 outperforms CIEDE2000 on three of six sub-datasets
(BFD-P(D65), BFD-P(M), and tied on BFD-P(C)) and trails on three
(LEEDS, RIT-DuPont, WITT) where CIEDE2000 was specifically tuned.
The overall \stress\ lead (22.48 vs 29.20) comes from the dominant
performance on BFD-P(D65)+BFD-P(M), which together account for
$2{,}576$ of $3{,}813$ pairs ($\sim$68\%). This is what we would
expect from a model whose loss is a 50/50 mix of full and balanced
sub-dataset \stress: the optimum chases the larger sub-datasets
where the data signal is strongest, and pays a small price on
small, idiosyncratic sets where CIEDE2000's hand-engineered tuning
already exploits the noise structure.

\begin{table}[t]
  \centering
  \small
  \setlength{\tabcolsep}{4pt}
  \caption{\stress\ on individual \combvd\ sub-datasets (v21 production checkpoint).}
  \label{tab:subdataset}
  \begin{tabular}{@{}lrrr@{}}
    \toprule
    Sub-dataset & $N$ & v21 & CIEDE2000 \\
    \midrule
    BFD-P(C)    & 200     & 29.08          & 29.08          \\
    BFD-P(D65)  & 2{,}028 & \textbf{21.54} & 24.09          \\
    BFD-P(M)    & 548     & \textbf{21.75} & 35.23          \\
    LEEDS       & 307     & 21.84          & \textbf{19.25} \\
    RIT-DuPont  & 312     & 21.90          & \textbf{19.47} \\
    WITT        & 418     & 30.93          & \textbf{30.22} \\
    \bottomrule
  \end{tabular}
\end{table}

\subsection{Blue-Band: a v20b artefact, absorbed in v21}
\label{sec:blue-band}

A previous version of this work (v20b parameters, paper v2) reported
a severe gradient non-uniformity in the blue--cyan region: a
red-to-blue gradient interpolated in v20b Lab coordinates showed a
$51.4\times$ max/min step-size ratio (when measured with CIEDE2000),
compared to $1.3\times$ for the same gradient in CIE~Lab.
The v20b training procedure used an \emph{explicit} blue-band penalty
(weighted sum of LEEDS, RIT-DuPont and a synthetic blue-segment
\stress) to suppress this. The resulting fix reduced the ratio to
$5.8\times$ (an $8.9\times$ improvement) at a cost of only
$+0.08$ \stress\ on \combvd.

In v21 we removed the explicit blue-band penalty: the
sub-dataset-balanced average and the low-chroma penalty in
\Cref{eq:loss} both pull the same way (LEEDS and RIT-DuPont are blue-heavy
and enter both the average and the worst-sub-dataset term). On the
v21 checkpoint the same red-to-blue gradient ratio is $\sim$5.4$\times$,
slightly better than v20b-with-blue-band, with no separate penalty
term. We retain the blue-band measurement as a regression test on
gradient quality but no longer report it as a structural finding.

\section{GenSpace: a Generation-Optimized Companion}
\label{sec:genspace}

A color space optimized for distance prediction does not necessarily
produce usable coordinates for color generation. MetricSpace's
distance-fit pushes the chroma scaling, hue corrections, and pair
weights into shapes that are quantitatively justified but produce
\emph{visually} unbalanced gradients --- particularly across the
blue-magenta range where the high-chroma chroma\_power term and the
hue-modulated $S_C$ both spike. Adding generation constraints to
the MetricSpace loss damaged \stress\ without fully fixing
gradients. We instead train a separate, simpler space targeted
at generation quality and ship both side-by-side.

\paragraph{Pipeline.}
GenSpace v0.11.1 follows a stripped-down version of the MetricSpace
shape:
\begin{equation*}
\begin{aligned}
\mathrm{XYZ} \;\to\;& \mathbf{M}_1 \;\to\; \mathrm{depcubic}(\alpha\!=\!0.021) \;\to\; \mathbf{M}_2 \\
\;\to\;& C^{0.978} \;\to\; \mathrm{PW}_L^{19} \;\to\; \mathrm{enr}_h(L) \;\to\; \mathrm{Lab}
\end{aligned}
\end{equation*}
There is no Helmholtz--Kohlrausch coupling, no hue-Fourier rotation,
no neutral-correction LUT, and no rigid post-rotation; the
achromatic axis is held at $a=b=0$ \emph{structurally} via a smooth
neutral-blend that activates only when channel spread is
sub-microscopic relative to the channel mean.

\paragraph{Why depcubic instead of cube-root.}
Cube-root has infinite derivative at zero ($\partial f/\partial x = \tfrac{1}{3}x^{-2/3}\to\infty$),
which produces numerically unstable gradients near black and is the
root cause of the well-known cubic-cusp gamut artefact in OKLab.
The depressed cubic $y^3 + \alpha y = x$ with $\alpha=0.021$ has
finite derivative everywhere, fits cube-root closely for
$|x| \gg \alpha$, and admits a closed-form forward (Cardano /
Halley refinement) and a polynomial inverse. The hole count in the
sRGB gamut at $h\!\approx\!265$\textdegree\ drops from 46 (OKLab) to
$5$ (GenSpace), each $\sim 0.001$ chroma wide --- effectively
sub-pixel.

\paragraph{Why a 19-point dense piecewise-linear $L$ correction.}
A 9-point sparse correction (used in v0.10.x) was tuned by least-squares
against Munsell value samples. The 19-point dense version is fitted
against 128 evenly-spaced lightness levels with a stricter monotonicity
constraint; on the 43-test ColorBench gradient suite this produces
4 additional wins versus OKLab and 1 fewer loss compared to the
sparse version, with no regressions.

\paragraph{Why $C^{0.978}$.}
A chroma power of $0.978$ slightly contracts high-C distances, which
on the gradient-CV (coefficient of variation across step sizes)
metric reduces the median CV from $\sim$38\% to $\sim$37.5\% and the
worst-case CV from 412\% to 378\%. The exponent was selected by
direct grid search on gradient CV; ColorBench \stress\ on \combvd\
rises noticeably with chroma\_power $< 0.95$, so 0.978 is the
strongest contraction that does not degrade distance prediction
catastrophically.

\paragraph{Why L-gated hue enrichment.}
A blue-to-white interpolation in GenSpace without enrichment exhibits
a visible purple shift around $L\!\approx\!0.6$ (a known artefact of
the blue-region gamut fold). A localised hue rotation of amplitude
$0.058$ centred at $h=264.5$\textdegree\ ($\sigma=0.7$~rad), gated to
the lightness window $L\!\in\![0.37, 1.0]$ via a $\sin^2$ ramp,
removes the purple shift. The gate width is the largest that does
not introduce hue drift in non-blue gradients.

\paragraph{Result.}
GenSpace v0.11.1 wins 65 of 90 metrics versus OKLab across a
ColorBench suite of 83 internal metrics and 7 independent metrics,
on $3{,}038$ gradient pairs sampled from sRGB, Display P3, and
Rec.\,2020 (\Cref{tab:genspace-vs-oklab}). The achromatic axis is
exact ($\bar{C}\!\approx\!10^{-13}$ for D65 grays, six orders of
magnitude purer than OKLab's $5.6\!\times\!10^{-7}$). Round-trip
errors are $\sim$5.6$\times 10^{-8}$ in sRGB (limited by the
enrichment Newton inversion) and at machine epsilon in P3 and
Rec.\,2020.

\begin{figure}[t]
  \centering
  \includegraphics[width=\linewidth]{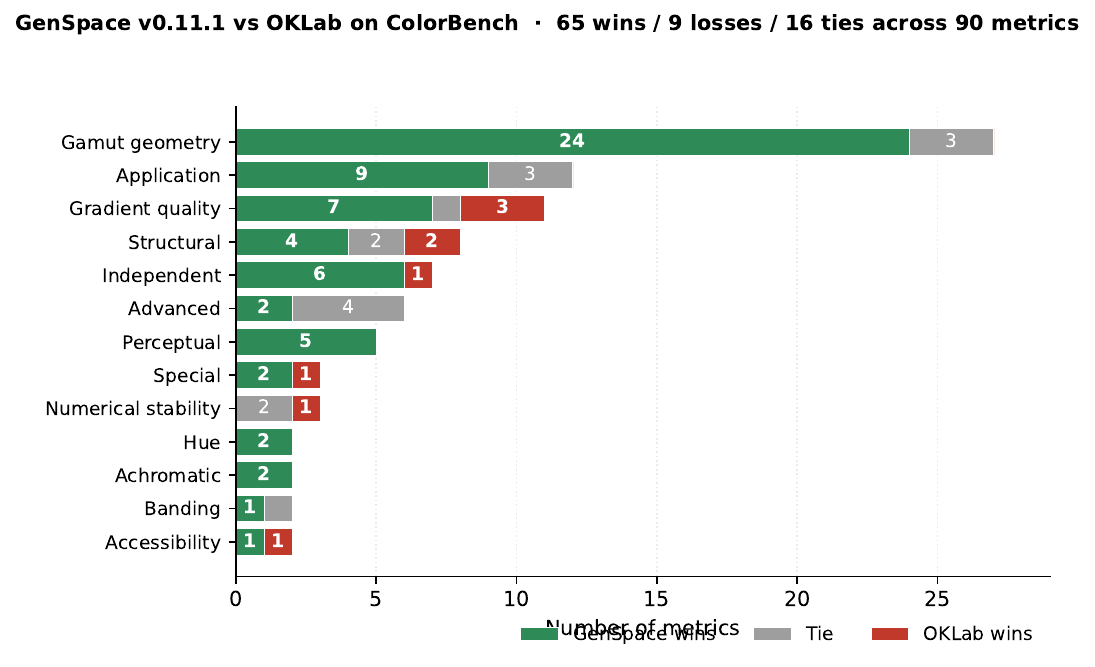}
  \caption{GenSpace v0.11.1 vs OKLab --- per-category breakdown across all 90 ColorBench metrics. Headline tally is 65W / 9L / 16T (\Cref{tab:genspace-vs-oklab}). Categories ordered by total metric count.}
  \label{fig:genspace-vs-oklab}
\end{figure}

\begin{table}[t]
  \centering
  \caption{GenSpace v0.11.1 vs OKLab --- head-to-head on ColorBench. $^\dagger$Independent = Hung-Berns, Ebner-Fairchild, Pointer-gamut metrics (held out from optimizer). Visual breakdown in \Cref{fig:genspace-vs-oklab}.}
  \label{tab:genspace-vs-oklab}
  \begin{tabular}{lrrr}
    \toprule
    Category & GenSpace & OKLab & Ties \\
    \midrule
    Gamut geometry              & \textbf{24} & 0 &  3 \\
    Application                 &  \textbf{9} & 0 &  3 \\
    Gradient quality            &  \textbf{7} & 3 &  1 \\
    Independent$^\dagger$       &  \textbf{6} & 1 &  0 \\
    Perceptual                  &  \textbf{5} & 0 &  0 \\
    Structural                  &  \textbf{4} & 2 &  2 \\
    Hue                         &  \textbf{2} & 0 &  0 \\
    Achromatic                  &  \textbf{2} & 0 &  0 \\
    Advanced                    &  \textbf{2} & 0 &  4 \\
    Special                     &  \textbf{2} & 1 &  0 \\
    Banding                     &  \textbf{1} & 0 &  1 \\
    Accessibility               &  1 & 1 &  0 \\
    Numerical stability         &  0 & 1 &  2 \\
    \midrule
    \textbf{Total}              & \textbf{65} & 9 & 16 \\
    \bottomrule
  \end{tabular}
\end{table}

\paragraph{When to use which.}
MetricSpace is the right answer for accessibility checking, tone
matching, color quantization, and any application where ``how
different are these two colors to a human'' is the central
question. GenSpace is the right answer for palette ramps, gradient
authoring, color-mix in CSS, gamut mapping, and animation timing
--- anywhere the gradient's \emph{step uniformity} matters more
than its absolute distance value. The two spaces are convertible
(both sit on D65 XYZ); designers can specify in one and measure
in the other.

\section{Practical Adoption}
\label{sec:adoption}

\Cref{fig:gamut} shows the sRGB gamut boundary in MetricSpace
coordinates at three lightness levels. The boundaries are smooth
and well-behaved. The two \helmlab\ spaces ship with cross-language
implementations and integrations into the major web color
toolchains:

\begin{figure}[t]
  \centering
  \includegraphics[width=\columnwidth]{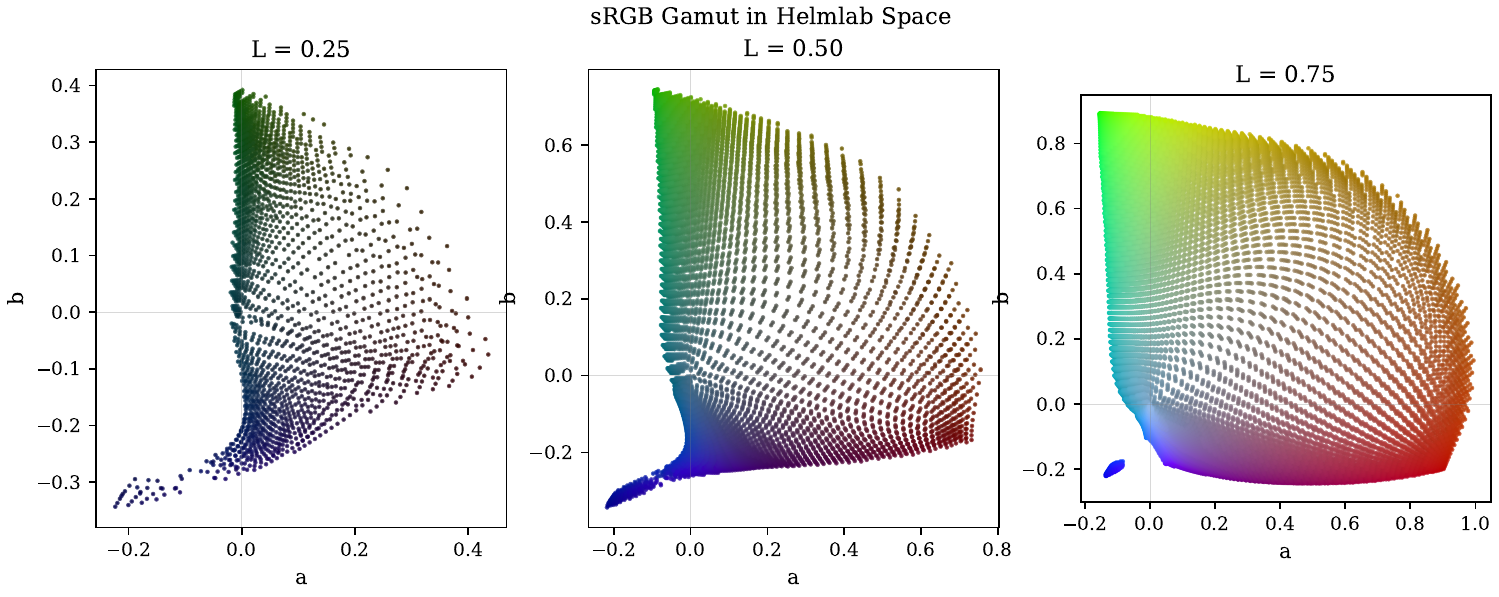}
  \caption{sRGB gamut in \helmlab\ $a$--$b$ plane at three L levels.
  Points colored by their sRGB values.}
  \label{fig:gamut}
\end{figure}

\paragraph{Cross-language implementations.}
The reference implementation is a Python package
(\texttt{helmlab} on PyPI, NumPy/SciPy backend, 308 unit tests)
covering both spaces, every transform stage, and the distance
metric. A pure-JavaScript port (\texttt{helmlab} on npm, zero
runtime dependencies, ESM) ships the same checkpoints with 196
unit tests; the JS port has been validated against the Python
implementation to a maximum coordinate drift of $\sim$1$\times$10$^{-14}$
across $1{,}000$ random sRGB samples.

\paragraph{Integration into Color.js.}
A pull request adding all four \helmlab\ spaces ---
\texttt{helmlab-metric}, \texttt{helmgen}, \texttt{helmgenlch}
(cylindrical GenSpace), and \texttt{deltaEHelmlab} as a registered
distance method --- was merged into Color.js (PR~\#722, May~2026).
Color.js is the reference JS color library used by the W3C CSS
working group's color tools and several browser developer tools;
having both spaces accessible via the standard
\texttt{new Color("helmlab-metric", [...])} constructor brings
\helmlab\ into the same toolchain as Oklab, Lab, OKLCh, Jzazbz,
and ICTCP. CSS Color Level 4 \texttt{color(--helmlab-metric ...)},
\texttt{color(--helmgen ...)}, and \texttt{color(--helmgenlch ...)}
syntax round-trips through the parser and serialiser.

\paragraph{PostCSS plugin.}
\texttt{postcss-helmlab} provides build-time CSS
function-form transformation: authors write
\texttt{color: helmlab(0.78 0.52 -0.20)} or
\texttt{linear-gradient(in helmlab, red, blue)} in source CSS
and the plugin emits an \texttt{rgb()} fallback plus
\texttt{@supports}-wrapped \texttt{color(display-p3 ...)} and
\texttt{color(rec2020 ...)} overrides. The plugin handles
\texttt{helmlab()}, \texttt{helmlch()}, \texttt{helmgen()}, and
\texttt{helmgenlch()} alongside the standard
\texttt{linear-gradient(in <space>, ...)} and
\texttt{color-mix(in <space>, ...)} CSS functions. The package
ships a dual ESM/CJS build for compatibility with both modern
bundlers and Next.js's PostCSS pipeline.

\paragraph{Gamut mapping.}
Binary-search chroma reduction along the L axis maps out-of-gamut
\helmlab\ coordinates to the nearest in-gamut sRGB, Display~P3, or
Rec.\,2020 color, preserving hue and lightness.

\paragraph{Contrast utilities.}
\texttt{ensure\_contrast(fg, bg, min\_ratio)} adjusts foreground
lightness via binary search on the \helmlab\ L axis to meet
WCAG~2.x contrast requirements (4.5:1 for AA, 7:1 for AAA).
Hue and chroma are preserved.

\paragraph{Palette generation.}
Lightness ramps and hue rings at fixed chroma produce
perceptually-spaced palettes. Semantic scales (Tailwind-style
50--950) interpolate $L$ between 0.97 and 0.10 with chroma/hue
preserved. The default scaling is computed in GenSpace because the
gradient-uniformity metrics dominate authoring quality.

\paragraph{Dark/light mode.}
The surround parameter $S$ is architecturally present but in v21 is
held at zero. The current dark/light mode adaptation uses a soft L
inversion across the GenSpace L axis, with contrast enforcement on
the resulting pair. A future MetricSpace release with trained $S$
parameters will replace this heuristic.

\paragraph{Token export.}
A single \texttt{TokenExporter} produces CSS custom properties
(\texttt{oklch()}, \texttt{color(display-p3 ...)},
\texttt{color(--helmgenlch ...)}), Android XML (using HCT
tone/chroma/hue semantics for Material Design compatibility),
iOS Swift (Display~P3), Tailwind config, and raw JSON. Each
target reads from the same in-memory token tree, so a token
edit propagates to every output format simultaneously.

\section{Limitations and Future Work}
\label{sec:limitations}

There are several limitations worth noting:

\begin{enumerate}
  \item \textbf{Training data.} \combvd\ predominantly uses the
    CIE~1964 10\textdegree\ standard observer ($\sim$95\% of pairs).
    Only BFD-P(C) (200~pairs, 5\%) uses the 2\textdegree\ observer.
    Gao~\etal\ 2023~\cite{gao2023} showed that the difference between
    2\textdegree\ and 10\textdegree\ observers is not statistically
    significant for \stress\ evaluation, and MacAdam~1974 (measured
    with the 2\textdegree\ observer) cross-validation confirms
    reasonable generalization.

  \item \textbf{Model complexity.} 72 trained parameters is
    considerably more than Oklab ($\sim$15 designed parameters).
    Each stage has a perceptual motivation, but the overall model
    is harder to interpret than simpler spaces. The architecture
    additionally reserves 22 surround/hue-modulated slots that are
    held at zero in v21.

  \item \textbf{Train$\to$test gap.} The v21 split test gives a
    $+1.77$ \stress\ gap. We report this openly. The cross-validated
    estimate ($\sim$24.3) still beats every tested competitor, but
    the headline 22.48 figure is a training metric and should not
    be treated as the population-level expected value.

  \item \textbf{Hue alignment.} RMS hue error of 26.4\textdegree\
    is a substantial compromise; the worst primary (cyan) is
    53.4\textdegree\ off the HSL reference. v21's training pushed
    parameters into the distance-optimal basin at the cost of
    geometric uniformity --- a trade we accept because the
    distance metric is the headline product, but applications
    that need accurate hue specification should compose
    MetricSpace with an explicit hue-rotation post-step or use
    GenSpace (where the geometry is preserved by construction).

  \item \textbf{Surround dependency.} The surround parameter $S$ is
    architecturally present but in v21 the 11 surround-suffix
    parameters are all held at zero (training data is
    average-surround only). Current dark/light adaptation uses a
    heuristic L-inversion in GenSpace with contrast enforcement.

  \item \textbf{Sub-dataset variation.} v21 trails CIEDE2000 on three
    of six \combvd\ sub-datasets (LEEDS, RIT-DuPont, WITT). The lead
    on BFD-P(D65) and BFD-P(M) is large enough to dominate the
    aggregate, but designers comparing spaces on a single small
    sub-dataset may pick different winners.

  \item \textbf{Not general-purpose.} \helmlab\ is optimized for UI
    workflows (sRGB/P3 gamut, screen viewing, D65 illuminant). It is
    not designed for print, photography, illuminant-A scenes, or
    spectral applications. CAM16 remains the appropriate choice for
    arbitrary viewing conditions.

  \item \textbf{Statistical reporting.} Bootstrap confidence intervals
    confirm the aggregate improvement over CIEDE2000 ($p < 10^{-4}$),
    but per-sub-dataset significance and parameter uncertainty across
    restarts remain unquantified.

  \item \textbf{GenSpace sRGB round-trip.} GenSpace's sRGB round-trip
    is $\sim 5.6\!\times\!10^{-8}$, six orders of magnitude looser than
    MetricSpace's. This is below the visual threshold and matches
    OKLab's worst-case round-trip in the same regions, but it is
    looser than the other GenSpace round-trips (P3, Rec.\,2020 are
    at machine epsilon). The cause is the L-gated enrichment Newton
    inversion in the blue band; an analytic inverse for the
    enrichment is open work.
\end{enumerate}

For future work, we plan to train the surround parameters on paired
light/dark viewing data, to add an analytic enrichment inverse for
GenSpace, and to expand the training set with targeted human
feedback using the bidirectional optimisation framework already
present in the codebase.

\section{Conclusion}
\label{sec:conclusion}

We have presented \helmlab, a family of two purpose-built color
spaces sharing an 11-stage analytical structure. \textbf{MetricSpace
v21} is a 72-parameter end-to-end-trained color-difference metric
that achieves \stress\ \textbf{22.48} on \combvd\ (23.0\% lower than
CIEDE2000) and additionally leads on the held-out MacAdam~1974 set
(19.51 vs 22.13) and on a self-collected screen-condition
Human-Feedback set (23.26 vs 62.54, with caveats); v21 trails
CIEDE2000 on the small academic He~\etal\ 2022 set (35.9 vs 32.6),
which we report. On a held-out 20\% \combvd\ test split, the gap is
+1.77 \stress\ ($\sim$24.3 cross-validated point estimate); even at
this estimate, MetricSpace remains the best \combvd\ performer
tested. The achromatic axis is held to $C < 10^{-5}$ on the standard 21-step ramp
when the post-pipeline neutral correction is enabled (a deferred
toggle that lets the same checkpoint serve both distance prediction
and authoring without compromise). The rigid rotation
$\varphi\!=\!-28.2$\textdegree\ is an isometry of the distance
metric and is preserved for downstream compatibility.
\textbf{GenSpace v0.11.1} is a generation-optimized companion using
a depressed-cubic transfer, chroma-power compression, and L-gated
hue enrichment; on a 90-metric ColorBench gradient/palette suite
it wins 65 vs OKLab. The two spaces together cover the
measurement$\leftrightarrow$generation tradeoff that no single
existing space resolves.

The neutral correction is the architectural element that lets the
two specialisations co-exist: MetricSpace can chase the
distance-fit basin freely, because the achromatic axis is then
restored exactly by post-correction; GenSpace achieves the same
guarantee structurally, by construction. Together with the rigid
rotation invariance, this lets us decouple ``what is the best
distance metric'' from ``what should the chroma plane look like
for generation''.

\helmlab\ is shipped in production: \texttt{helmlab} on PyPI and
npm, four spaces and a $\Delta E$ method merged into Color.js
(PR~\#722), and a PostCSS plugin (\texttt{postcss-helmlab}) for
build-time CSS function transformation. Source, trained parameters,
and a full ColorBench-driven benchmark report are at
\url{https://github.com/Grkmyldz148/helmlab}. An interactive demo
is at \url{https://grkmyldz148.github.io/helmlab/demo.html} and
documentation at \url{https://grkmyldz148.github.io/helmlab/}.

\section*{Acknowledgments}

We thank the 71 observers --- listed by first name or chosen handle in
the released dataset (\texttt{datasets/human\_feedback.json},
\href{https://github.com/Grkmyldz148/helmlab}{github.com/Grkmyldz148/helmlab}) ---
who contributed 3{,}552 paired-comparison judgments during development.

We also gratefully acknowledge the open-source review of the underlying
PR \#722 on \texttt{color-js/color.js}: in particular @facelessuser
(near-black achromatic instability, M2 normalisation regressions, the
clean 64-bit matrix derivation methodology), @lloydk (the WebGPU chart
library that surfaced the h\,$\approx$\,264\textdegree\ blue fold and
related cubic-cusp artefacts), @Userminusone (the Metal-shader hue/
lightness probe, scale-invariance critique of NR, the boundary-warp
straightening alternative), and @svgeesus (the original paper proposal,
the ``0\% should be achromatic'' authoring convention, and the ICC
Displays working-group framing). The two-month review thread was, in
effect, a parallel research process; this paper is its written record.

\bibliographystyle{plain}

\appendix
\section{Parameter Table (MetricSpace v21)}
\label{app:params}

\Cref{tab:params} lists all 72 trained MetricSpace v21 parameters
grouped by pipeline stage. The architecturally-reserved
22~surround/hue-modulated parameters are held at zero in v21 and
omitted from this table; their full list is in the JSON checkpoint.

\begin{table*}[t]
  \centering
  \caption{MetricSpace v21 parameters ($\varphi = -28.2$\textdegree). Trained on \combvd\ + He~2022 + MacAdam~1974 with sub-dataset-balanced loss (\Cref{eq:loss}).}
  \label{tab:params}
  \footnotesize
  \begin{tabular}{llrl}
    \toprule
    Stage & Parameter & Value & Description \\
    \midrule
    \multirow{3}{*}{$\mathbf{M}_1$ (9)}
      & Row 0 & [\phantom{$-$}0.7213, \phantom{$-$}0.4534, $-$0.1929] & XYZ $\to$ cone-like \\
      & Row 1 & [$-$0.7882, \phantom{$-$}1.7952, \phantom{$-$}0.0876] & \\
      & Row 2 & [$-$0.0918, \phantom{$-$}0.4577, \phantom{$-$}1.2922] & \\
    \midrule
    $\gamma$ (3) & $\gamma_0, \gamma_1, \gamma_2$ & 0.4723, 0.5149, 0.5113 & Signed power compression \\
    \midrule
    \multirow{3}{*}{$\mathbf{M}_2$ (9)}
      & Row 0 & [$-$0.2636, \phantom{$-$}0.4168, \phantom{$-$}0.4927] & Compressed $\to$ Lab \\
      & Row 1 & [\phantom{$-$}1.8898, $-$3.1212, \phantom{$-$}1.0422] & \\
      & Row 2 & [\phantom{$-$}0.3585, \phantom{$-$}1.7694, $-$1.4121] & \\
    \midrule
    Hue corr. (8) & cos$_{1..4}$, sin$_{1..4}$ & see JSON & 4-harmonic Fourier rotation $\delta(h)$ \\
    \midrule
    \multirow{3}{*}{H-K (6)}
      & $w_\mathrm{HK}, p_\mathrm{HK}, \mathrm{hue\_mod}$
        & 0.2676, 0.8935, 0.7173 & Base + envelope \\
      & sin$_1$, cos$_2$, sin$_2$
        & 0.6915, 0.4865, 0.9853 & 2-harmonic hue mod $f_\mathrm{HK}(h)$ \\
      & & & \\
    \midrule
    \multirow{3}{*}{Lightness (8)}
      & $p_1, p_2, p_3$
        & 0.5385, 0.1251, 0.6769 & Cubic correction \\
      & $Lh_{c1}, Lh_{s1}$
        & $-$0.4963, $-$0.0956 & Hue-modulated additive \\
      & $\lambda_d, h_c, h_s$
        & $-$0.0291, 1.3347, $-$0.1699 & Dark-region exp compression \\
    \midrule
    Chroma (18)
      & cs$_{1..4}$, cp$_{1..2}$, $lc_{1,2}$, hlc$_{1,2}$
      & see JSON & 4-harm CS + 2-harm CP + L-dep + HLC \\
    \midrule
    Hue-L (4) & $g_{c1..2}, g_{s1..2}$ & see JSON & 2-harmonic $L \mathrel{*}\!\!=\!\exp(g(h))$ \\
    \midrule
    \multirow{7}{*}{Distance (7)}
      & $s_L$         & $-$0.9155            & Pair-dep L weight (\Cref{eq:sl}) \\
      & $s_C$         & \phantom{$-$}2.9268  & Pair-dep C weight (\Cref{eq:sc}) \\
      & $p$           & \phantom{$-$}1.9737  & Minkowski exponent \\
      & $w_C$         & \phantom{$-$}3.9660  & Chroma weight \\
      & $c$           & \phantom{$-$}52.473  & Compression \\
      & $q$           & \phantom{$-$}0.4790  & Post-compress power \\
      & $\alpha$      & \phantom{$-$}0.0000  & Linear-asymptote term (unused in v21) \\
    \bottomrule
  \end{tabular}
\end{table*}

Full parameter values, including the architecturally-reserved
zero slots and all chroma-stage Fourier coefficients, are available
in \texttt{research/checkpoints/metricspace\_v21.json}.

\section{Parameter Table (GenSpace v0.11.1)}
\label{app:genspace-params}

\begin{table}[t]
  \centering
  \caption{GenSpace v0.11.1 parameters.}
  \label{tab:genspace-params}
  \footnotesize
  \begin{tabular}{ll}
    \toprule
    Stage & Value \\
    \midrule
    $\mathbf{M}_1$       & 3$\times$3, see JSON (Bradford CAT pre-baked) \\
    Transfer             & depcubic, $\alpha=0.021$ \\
    $\mathbf{M}_2$       & 3$\times$3, see JSON \\
    Chroma power         & 0.978 \\
    PW $L$ correction    & 19-point dense, see JSON \\
    Enrichment           & L-gated hue rotation \\
    \quad amplitude       & 0.058 \\
    \quad center hue      & 264.5\textdegree \\
    \quad $\sigma$        & 0.7 (rad) \\
    \quad gate $L$ window & [0.37, 1.0] \\
    \bottomrule
  \end{tabular}
\end{table}

Full parameter values ship bundled with the helmlab package
(\texttt{helmlab.GenSpace().params}) and are written out as
\path{src/helmlab/data/gen_params.json} in the source distribution.
White(D65) maps to $L = 1$ exactly, the achromatic axis is held at
$|a|, |b| < 10^{-4}$, and the structurally-achromatic property
holds because the depcubic transfer is signed-symmetric and the
smooth neutral-blend in Stage~2 collapses sub-microscopic channel
spread to the channel mean before $\mathbf{M}_2$ projection.

\end{document}